\documentclass[twocolumn,showpacs]{revtex4}
\usepackage{graphicx}

\newcommand{\bastar}{\begin{eqnarray*}}
\newcommand{\eastar}{\end{eqnarray*}}
\newskip\humongous \humongous=0pt plus 1000pt minus 1000pt

\newif\ifdtup

\relax
\newcommand{\be}{\begin{equation}}
\newcommand{\ee}{\end{equation}}
\newcommand{\bea}{\begin{eqnarray}}
\newcommand{\eea}{\end{eqnarray}}

\newcommand{\n}{\hat n}

\newcommand{\hn}{{\hat n}}

\newcommand{\dfrac}{\displaystyle\frac}
\newcommand{\ba}{\begin{array}}
\newcommand{\ea}{\end{array}}

\newcommand{\nn}{\nonumber}

\newcommand{\bD}{\bar D}
\begin{document}
\title{Gauge Invariance and Stability of SNO vacuum in QCD}
\bigskip

\author{Y. M. Cho}
\email{ymcho@yongmin.snu.ac.kr}
\affiliation{C.N.Yang Institute for Theoretical Physics,
State University of New York, Stony Brook, \\
New York 11794, USA \\
and \\
School of Physics,
College of Natural Sciences,
Seoul National University, \\
Seoul 151-747, Korea }
\begin{abstract}
We point out a critical defect in the calculation of the functional
determinant of the gluon loop in the Savvidy-Nielsen-Olesen (SNO) 
effective action.
We prove that the gauge invariance naturally exclude the unstable
tachyonic modes from the gluon loop integral. This
guarantees the stability of the magnetic condensation in QCD.
\end{abstract}
\maketitle

The confinement problem in quantum chromodynamics (QCD)
is probably one of the most challenging problems
in theoretical physics. It has
long been argued that the confinement in QCD
can be triggered by the monopole condensation \cite{nambu,cho1}.
Indeed, if one assumes monopole condensation, one can easily argue
that the ensuing dual Meissner effect could guarantee the confinement
of color \cite{hooft,cho2}. But it has been extremely difficult
to prove the monopole condensation in QCD. 

A natural way to establish the monopole condensation in QCD
is to demonstrate that the quantum fluctuation triggers
a phase transition through the dimensional transmutation
known as the Coleman-Weinberg mechanism \cite{cole}.
To prove the monopole condensation, one need to demonstrate
such a phase transition in QCD. There have been 
many attempts to do so with the one-loop effective action 
of QCD \cite{savv,niel,ditt}. Savvidy has first calculated
the effective action of $SU(2)$ QCD in the presence of an {\it ad hoc}
color magnetic background, and has almost ``proved''
the magnetic condensation. In particular, he
showed that the quantum effective potential obtained
from the real part of the one-loop effective action
has the minimum at a non-vanishing magnetic background \cite{savv}.
Unfortunately, the calculation repeated by Nielsen and
Olesen showed that the effective action
contains an extra imaginary part which 
destablizes the magnetic condensation \cite{niel}.
This instability of the ``Savvidy-Nielsen-Olesen (SNO)
vacuum'' has destroyed the hope to establish the monopole condensation
in QCD with the effective action \cite{ditt}.

Recently, however, there has been a new attempt to calculate
the one-loop effective action of QCD
with a gauge independent separation
of the non-Abelian monopole background from
the quantum field \cite{cho3,cho4}. Remarkably, in this
calculation the effective action has been shown to
produce no imaginary part in the presence of
the monopole background, but a negative imaginary part
in the presence
of the pure color electric background. This implies that in QCD
the non-Abelian monopole background produces a stable monopole
condensation, but the color electric background
becomes unstable by generating a pair annhilation of
the valence gluon.
The new result sharply contradicts with the earlier results,
in particular on the stability of the monopole condensation.
This has resurrected the old controversy on the stability of monopole
condensation.

To resolve the controversy it is important to understand
the origin of the instability of the SNO vacuum.
The energy of a charged vector field
moving around a constant magnetic field depends on the
spin orientation of the vector field, and when the
spin is anti-parallel to the magnetic field, the zeroth
Landau level has a negative energy.
Because of this the functional determinant of the gluon loop
in the SNO magnetic background necessarily contains
negative eigenvalues which create a severe infra-red divergence
in the effective action. And, when one regularizes
this divergence with the $\zeta$-function regularization,
one obtains the well-known imaginary
component in the effective action which destablizes
the magnetic condensation \cite{niel}.
This tells that the instability of the SNO vacuum originates
from the the negative eigenvalues of the functional determinant.
Since the origin of the negative eigenvalues is so obvious,
the instability of the SNO vacuum has become the prevailing
view \cite{niel,ditt}.

\begin{figure*}
\includegraphics{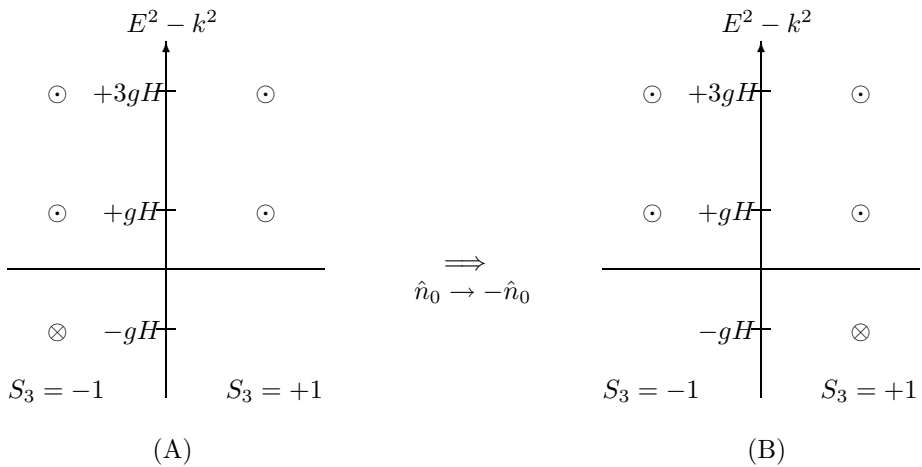}
\caption{\label{Fig. 1} The eigenvalues of the functional
determinant of the gluon loop. When the gluon spin is
anti-parallel to the magnetic field ($S_3=-1$), the ground state
(with $n=0$) becomes tachyonic when $k^2<gH$. Notice, however,
that under the color reflection of $\hn_0$ to $-\hn_0$
$H$ changes to $-H$ so that the eigenvalues change from (A) 
to (B). This shows that the spin polarization direction 
of gluon is a gauge artifact. This excludes the tachyons from 
the functional determinant.}
\end{figure*}

This view, however, is not without defect.
To see this notice that the eigenfuctions corresponding to
the negative eigenvalues describes tachyons which 
are unphysical. This implies that
one should exclude these tachyons
in the calculation of the effective action, unless
one wants to allow the violation of causality in QCD.
Unfortunately the standard $\zeta$-function regularization fails
to remove the contribution of the tachyonic eigenstates
because it is insensitive to causality. On the other hand,
if we adopt the infra-red regularization which respects the
causality, the resulting effective action has no
imaginary part \cite{cho3,cho4}. But since the $\zeta$-function
regularization has worked so
well in quantum field theory, there seems no compelling reason
why it should not work in QCD. So we need to find an independent
argument which can remove the negative eigenvalues from 
the functional determinant.

{\it The purpose of this paper is to show that a proper
implementation of the gauge invariance in the calculation of
the functional determinant of the gluon loop excludes
the unstable tachyonic modes,
and thus naturally restore the stability of the magnetic background.
This tells that it is
the incorrect calculation of the functional determinant,
not the $\zeta$-function regularization, which causes
the instability of the SNO vacuum}.
This means that tachyons should not have been
there in the first place. They were there to create
a mirage, not physical states. 

In the old approach Savvidy starts from the SNO
background which is not gauge invariant \cite{savv,niel}.
Because of this the functional determinant of the gluon loop
contains the tachyonic eigenstates when the gluon spin is
anti-parallel to the magnetic field. To cure this defect
Nielsen and Olesen has introduced the gauge invariant
``Copenhagen vacuum'' \cite{niel}. Although conceptually appealing,
however, this Copenhagen vacuum was not so useful in proving
the stability of the monopole condensation. 
In the following we show that the gauge invariant functional 
determinant should not depend on the spin polarization of 
the gluon. This tells that, if we impose the gauge invariance 
properly, the instability of the SNO background should disappear.

\begin{figure*}
\includegraphics{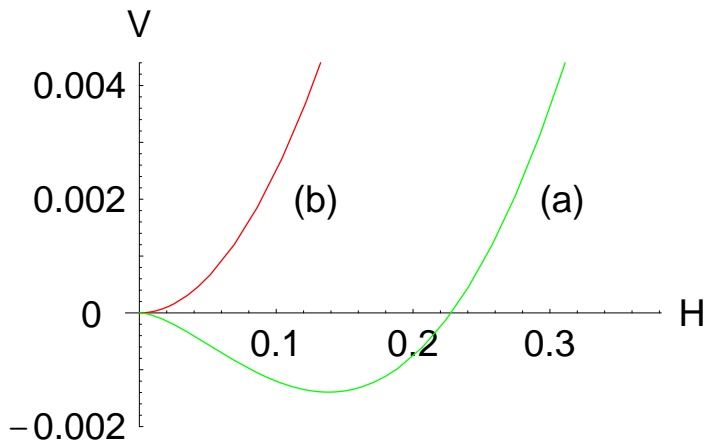}
\caption{\label{Fig. 2} The effective potential of SU(2)
QCD in the pure magnetic background.
Here (a) is the effective potential and (b) is the classical potential.}
\end{figure*}

To obtain the one-loop effective action one must divide
the gauge potential $\vec A_\mu$ into the slow-varying
classical background $\vec B_\mu$ and the fluctuating quantum
part $\vec Q_\mu$,
\bea
\vec A_\mu = \vec B_\mu + \vec Q_\mu,
\label{d}
\eea
and integrate out the quantum part. The gluon loop and 
the ghost loop integrals give the following functional 
deteminants \cite{savv,niel,cho3,cho4}
\bea
&{\rm Det}^{-\frac{1}{2}} K_{\mu \nu}^{ab} =
{\rm Det}^{-\frac{1}{2}} \Big(-g_{\mu \nu}
\bD^2_{ab}- 2g \epsilon_{abc} G_{\mu \nu}^c \Big),\nn \\
& {\rm Det} M_{ab} = {\rm Det} \Big(-\bD^2_{ab} \Big), 
\label{fd}
\eea
where $\bD_\mu$ and $\vec G_{\mu\nu}$ are the covariant 
derivative and field strength of the background $\vec B_\mu$.
From this one has
\bea
\Delta S = \dfrac{i}{2} \ln {\rm Det} K - i \ln {\rm Det} M.
\label{ea0}
\eea
Savvidy has chosen a covariantly constant color magnetic field 
as the classical background \cite{savv,niel,ditt}
\bea
&\vec B_\mu = \dfrac{1}{2} H_{\mu\nu} x_\nu \n_0,
~~~~~\vec G_{\mu\nu} = H_{\mu\nu} \n_0, \nn\\
&\bar D_\mu \vec G_{\mu\nu} = 0,
\label{sb}
\eea
where $H_{\mu\nu}$ is a constant magnetic field and $\n_0$ is
a constant unit isovector.
The calculation of the functional
determinant (\ref{fd}) amounts to the calculation of 
the energy eigenvalues of a massless charged vector field 
(the valence gluon) in a constant external
magnetic field $H_{\mu\nu}$ \cite{niel}.
Choosing the direction of the magnetic field to be the $z$-direction,
one obtains the well-known eigenvalues
\bea
&E^2 = 2gH (n + \dfrac{1}{2}) + k^2 \pm 2gH, \nn\\ 
&H = H_{12},
\label{ev}
\eea
where $k$ is the momentum of 
the eigen-function in $z$-direction.
Notice that the $\pm$ signature correspond to the spin $S_3=\pm 1$
of the valence gluon. So, when $n=0$,
the eigen-functions with $S_3=-1$ have an imaginary energy when
$k^2<gH$, and thus become tachyons which violate
the causality. 

From (\ref{ev}) one obtains
\bea
&\Delta S = i\ln {\rm Det} \Big[(-\bD^2+2gH)(-\bD^2-2gH) \Big],
\label{fdhx}
\eea
so that \cite{niel,ditt}
\bea
&{\cal L}_{eff} = -\dfrac{H^2}{4}-\dfrac{11g^2}{96\pi^2} H^2 (\ln
\dfrac{gH}{\mu^2}-c) + i \dfrac {g^2} {16\pi} H^2, 
\label{snoea}
\eea
where $c$ is a regularization-dependent connstant.
This contains the well-known imaginary part which destablizes
the SNO vacuum \cite{niel,ditt}. 

Notice, however, that the background $\vec G_{\mu\nu}$ 
must be gauge covariant. So one can
change $\vec G_{\mu\nu}$ to $-\vec G_{\mu\nu}$, and thus 
$H_{\mu\nu}$ to $-H_{\mu\nu}$, by a gauge transformation 
(with the color reflection of 
$\hn_0$ to $-\hn_0$) so that they are gauge equivalent. 
{\it This tells that the polarization direction of the magnetic 
background is a gauge artifact.
Furthermore, under this gauge transformation
the eigenvalues of $S_3=+1$ ($S_3=-1$) shift nagatively
(positively) by a factor $2gH$. And obviously only the eigenvalues
which are invariant under this transformation should qualify 
to be gauge invariant. This means that the gauge invariant
eigenstates are those
which are independent of the spin orientation
of the valence gluon which appear in both $S_3=+1$
and $S_3=-1$ simultaneously. In particular, this tells that 
the tachyonic eigenstates are not gauge invariant.} 
This is shown schematically in Fig. 1,
where (A) transforms to (B) (and vise versa) under the color
reflection. This tells that one must exclude
the tachyons in one's calculation of
the effective action.

If one does so,
the effective action (\ref{fdhx}) changes to
\bea
&\Delta S = i\ln {\rm Det} \Big[(-\bD^2+2gH)(-\bD^2+2gH) \Big],
\label{fdho}
\eea
which has no infra-red divergence at all.
This precludes the necessity to make any infra-red regularization.
From this we have \cite{cho3,cho4}
\bea
&{\cal L}_{eff} = -\dfrac{H^2}{4} -\dfrac{11g^2}{96\pi^2} H^2 (\ln
\dfrac{gH}{\mu^2}-c),
\label{eaho}
\eea
which clearly has no imaginary part.

A simple way to understand the above result is to remember 
that the effective action is nothing but the vacuum to vacuum
amplitude in the presence of the classical background,
\bea
&\exp \Big[i S_{eff} (\vec B_\mu)\Big]
= <\Omega_+|~\Omega_-> \Big|_{\vec B_\mu} \nn\\
&= \dfrac{}{} \sum_{|n_i>}<\Omega_+|~n_i>
<n_i~|~\Omega_-> \Big|_{\vec B_\mu},
\label{vtv}
\eea
where $|\Omega>$ is the vacuum and $|n_i>$ is
a complete set of orthonormal states
of QCD. In this view the gluon loop integral corresponds to 
the summation of the complete set of states.
And obviously the complete set should not include
the tachyons, unless one wants to assert that the physical 
spectrum of QCD must contain the unphysical states
which violate the causality and the gauge invariance.
This justifies the exclusion of the tachyons
in the calculation of the functional determinant.

The effective action (\ref{eaho}) generates the much desired
dimensional transmutation in QCD \cite{cho3,cho4}.
To demonstrate this notice that the effective action 
provides the following effective potential
\bea
V=\dfrac{H^2}{4}
\Big[1+\dfrac{11 g^2}{24 \pi^2}(\ln\dfrac{gH}{\mu^2}-c)\Big].
\eea
So we define the running coupling $\bar g$ by \cite{savv,cho3}
\bea
\frac{\partial^2V}{\partial H^2}\Big|_{H=\bar \mu^2}
=\frac{1}{2}\frac{g^2}{\bar g^2}.
\eea
With the definition we find
\bea
\frac{1}{\bar g^2} =
\frac{1}{g^2}+\frac{11}{24 \pi^2}( \ln\frac{{g\bar\mu}^2}{\mu^2}
- c + \dfrac{3}{2}),
\eea
from which we obtain the following $\beta$-function 
\bea
\beta(\bar\mu)= \bar\mu \dfrac{\partial \bar g}{\partial \bar\mu}
= -\frac{11 \bar g^3}{24\pi^2}.
\eea
This is exactly the same $\beta$-function that one obtained
from the perturbative QCD to prove the asymptotic 
freedom \cite{wil}. 

In terms of the running coupling the renormalized potential 
is given by
\bea
V_{\rm ren}=\dfrac{H^2}{4}
\Big[1+\dfrac{11 \bar g^2}{24 \pi^2 }
(\ln\dfrac{H}{\bar\mu^2}-\dfrac{3}{2})\Big],
\eea
which generates a non-trivial local minimum at 
\bea
<H>=\dfrac{\bar \mu^2}{\bar g} \exp\Big(-\frac{24\pi^2}{11\bar g^2}+ 1\Big).
\eea
Notice that with ${\bar \alpha}_s = 1$ we have
\bea
\dfrac{<H>}{{\bar \mu}^2} = 0.13819... .
\eea
This is nothing but the desired magnetic condensation.
The corresponding effective potential is plotted in Fig. 2,
where we have assumed $\bar \alpha_s = 1$ and $~\bar \mu =1$.

Nielsen and Olesen have suggested that the existence
of the unstable tachyonic modes are closely related with
the asymptotic freedom in QCD \cite{niel}.  Our analysis tells
that this need not be true. Obviously our asymptotic freedom
follows from a stable monopole condensation.

To establish the monopole condensation in QCD
with the effective action has been extremely difficult
to attain. The central issue here
has been the stability of the monopole
condensation.  The earlier attempts
to prove the monopole condensation
have produced a negative result, because the SNO background 
is not gauge invariant \cite{savv,niel}.
In this paper we have shown that a proper implementation of
gauge invariance naturally restores the stability
of the magnetic condensation.

It is not surprising that the gauge invariance plays the crucial role
in the stability of the monopole condensation. From the beginning
the gauge invariance has been the main motivation for the confinement
in QCD. It is this gauge invariance which forbids 
colored objects from the physical spectrum of QCD.
This necessitates the confinement of color.
So it is only natural that the gauge invariance
assures the stability of the monopole condensation, and thus
the confinement of color.

Finally it must be emphasized that there are actually two ways 
to exclude the unphysical modes, when one calculates 
the functional determinant or when one makes the infra-red 
regularization. We have already shown how to do this when we make
the infra-red regularization which respects the 
causality \cite{cho3,cho4}.
In this paper we have shown how to do this when we calculate
the functional determinant properly implementing the gauge 
invariance. And they produce the same effective action.
It is really remarkable that two completely independent principles,
the causality and the gauge invariance, both endorse
the stability of the magnetic condensation in QCD. 

The detailed discussion of the above result 
will be published elsewhere \cite{cho5}.

{\bf Acknowledgements}

~~~We thank Professor S. Adler and Professor F. Dyson
for the fruitful discussions, and Professor C. N. Yang for
the encouragements. This work is supported in part by
the ABRL Program of Korea Science and Engineering Foundation
(R14-2003-012-01002-0) and by BK21 Project of Ministry of Education.


\begin{thebibliography}{99}
\bibitem{nambu}Y. Nambu,
Phys. Rev. {\bf D10}, 4262 (1974);
S. Mandelstam, Phys. Rep. {\bf 23C}, 245 (1976);
A. Polyakov, Nucl. Phys. {\bf B120}, 429 (1977).
\bibitem{cho1}Y. M. Cho, Phys. Rev. {\bf D21}, 1080 (1980);
Phys. Rev. {\bf D62}, 074009 (2000).
\bibitem{hooft} G. 't Hooft, Nucl. Phys. {\bf B190}, 455 (1981).
\bibitem{cho2}Y. M. Cho, Phys. Rev. Lett. {\bf 46}, 302 (1981);
Phys. Rev. {\bf D23}, 2415 (1981); W. S. Bae, Y. M. Cho,
and S. W. Kimm, Phys. Rev. {\bf D65}, 025005 (2002).
\bibitem{cole} S. Coleman and E. Weinberg, Phys. Rev. {\bf D7},
1888 (1973).
\bibitem{savv} G. K. Savvidy, Phys. Lett. {\bf B71}, 133 (1977).
\bibitem{niel} N. Nielsen and P. Olesen, Nucl. Phys. {\bf B144}, 485 (1978);
{\bf B160}, 380 (1979); C. Rajiadakos, Phys. Lett. {\bf B100}, 471 (1981).
\bibitem{ditt} A. Yildiz and P. Cox, Phys. Rev. {\bf D21}, 1095
(1980);
M. Claudson, A. Yilditz, and P. Cox, Phys. Rev. {\bf D22}, 2022
(1980);
W. Dittrich and M. Reuter, Phys. Lett. {\bf B128}, 321, (1983);
{\bf B144}, 99 (1984); C. Flory, Phys. Rev. {\bf D28}, 1425 (1983);
S. K. Blau, M. Visser, and A. Wipf, Int. J. Mod. Phys.
{\bf A6}, 5409 (1991); M. Reuter, M. G. Schmidt, and C. Schubert,
Ann. Phys. {\bf 259}, 313 (1997).
\bibitem{cho3} Y. M. Cho and D. G. Pak, Phys. Rev. {\bf D65},
074027 (2002); Y. M. Cho, H. W. Lee, and D. G. Pak,
Phys. Lett. {\bf B 525}, 347 (2002).
\bibitem{cho4} Y. M. Cho, M. L. Walker, and D. G. Pak, JHEP {\bf 05},
073 (2004); Y. M. Cho and M. L. Walker, hep-th/0206127.
\bibitem{wil} D. Gross and F. Wilczek, Phys. Rev. Lett. {\bf 26},
1343 (1973); H. Politzer, Phys. Rev. Lett. {\bf 26}, 1346 (1973).
\bibitem{cho5} Y. M. Cho, hep-th/0301013.
\end{thebibliography}
\end{document}